\documentclass[aps,prb,twocolumn]{revtex4}
\usepackage{epsfig}
\usepackage[dvipsnames,usenames]{color}
\usepackage{color}
\usepackage{amsmath}
\usepackage{amssymb}
\usepackage{hyperref}
\usepackage{graphicx}

\usepackage{wasysym}
\usepackage{times}
\usepackage{comment}
\usepackage{array}
\usepackage{multirow}
\usepackage{tabularx}
\usepackage{float}
\usepackage[utf8]{inputenc}
\usepackage[T1]{fontenc}
\usepackage{cancel}
\usepackage{ulem}

\emergencystretch=\maxdimen
\begin{document}
\title{Shifting superconducting dome of two-dimensional hole-doped Hubbard model by a dry Fermi sea: \\ enhanced d-wave pairing in the overdoped regime}
\author{Mi Jiang}
\affiliation{Institute of Theoretical and Applied Physics, Jiangsu Key Laboratory of Thin Films, School of Physical Science and Technology, Soochow University, Suzhou 215006, China}

\begin{abstract}
Motivated by the presence of an extremely small electron pocket in the electronic band structure of infinite-layer (IL) nickelate superconductors, we systematically explore the superconducting (SC) properties of a two-dimensional Hubbard model influenced by an additional conduction band with small occupancy (dry Fermi sea). Our dynamic cluster quantum Monte Carlo calculations demonstrated the unusual impact of this additional dry metallic layer, which shifts the SC dome of the correlated Hubbard layer towards the overdoped regime, namely the $d$-wave SC can be enhanced (suppressed) in the overdoped (underdoped) regime. Simultaneously, the pseudogap (PG) crossover line shifts to the same higher doping regime as well.
Our analysis revealed the interplay of the effective pairing interaction, pair-field susceptibility, and antiferromagnetic spin fluctuation in the underlying non-monotonic physics. We postulate on the physical picture of SC dome shift in terms of the Kondo-type electron-hole binding in the case of a dry additional conduction band. Our investigation provided valuable new insights on tuning the SC properties via external dry bands.
\end{abstract}

\maketitle

\section{Introduction}
Unconventional superconductivity (SC) such as cuprate superconductors are widely believed to mainly originate from the electron-electron interaction despite that there are accumulating evidence supporting the equally important role of additional attractive channels such as electron-phonon interaction~\cite{phonon1}, whose interplay with the pure electron-electron interaction deserves more systematic exploration. 
These investigation reconcile with the long standing question of how to enhance the SC transition temperature $T_c$. 
In the last decades, some proposals have suggested that the ``incipient'' bands, namely full (empty) bands slightly below (above) the Fermi energy, can significantly enhance $T_c$ owing to either interband pair-scattering channels~\cite{incipient1,incipient2} or contributing to the spin fluctuation~\cite{incipient3}; while other study pointed out that the incipient band is not beneficial for enhancing $T_c$~\cite{incipient4}.
Recently, it has been uncovered that the exchange of pairs between half-filled wide and narrow bands can boost $T_c$~\cite{Werner}.
Besides, it was proposed that $T_c$ could be raised via coupled planes with different doped hole concentrations such that the high pairing scale and large phase stiffness are derived from the underdoped and overdoped planes separately~\cite{Kivelson}. Further studies  have been extensively conducted by both perturbative~\cite{Berg2008} and many-body calculations~\cite{Wachtel2012,RTS2014,Maier2022}.

The search for new families of superconductors with $T_c$ even higher than the cuprates constitutes one of the major goals but demanding task. The recent exploration of SC in infinite-layer (IL) nickelates~\cite{2019Nature,Pr,La,LaSC,Nd6Ni5O12} started a new era by uncovering a long-awaited unconventional SC system similar to cuprate SC despite that the current available $T_c$ is much lower than their cuprate counterparts~\cite{Junjie_review,Botana_review,Arita_review}. Because the intrinsic difficulty of the synthesis of IL-nickelates limits their experimental investigation~\cite{synthesis,LSAT1,LSAT},
the smoking-gun evidence of the electronic structure revealed by various theoretical studies is still lacking. 
Most of these electronic structure calculations agree that one significant difference between the two classes of materials is the existence of a rather broad band crossing the Fermi energy for the nickelates, which is composed of a combination of orbitals including Nd-$5d_{xy}$, Nd-$6s$, O-$2p$, Ni-$3d_{z^2}$, Ni-$4s$, and interstitial states. Intriguingly, this broad band of a mixed orbital character forms small electron pockets at $\Gamma$ and A points, which suggests that the IL-nickelates have distinct low-energy physics from the cuprates~\cite{Botana_review,Arita_review,Held2022,Hanghuireview}.

The most straightforward way of treating this additional tiny electron pocket mainly originating from the rare-earth layer of the IL-nickelates is to employ a minimal two-orbital $d$-$s$ model~\cite{TPD_ds,MJ2022,Oles2022}, where $d$ orbital mimics Ni-$3d$ orbital while the effective interstitial $s$ orbital realizes the effects of rare-earth R-$5d$ orbitals. Although this simplified model cannot describe the detailed physics of the IL-nickelates, our previous investigation uncovered some signature of the unusual impact of $s$ orbital on the pairing tendency in the correlated $d$-orbital layer~\cite{MJ2022}.  
Motivated by these earlier exploration, here by employing large scale dynamic cluster quantum Monte Carlo calculations, we will demonstrate that this ``dry'' metallic layer, at moderate $d$-$s$ hybridization, shifts the $d$-wave SC dome in the phase diagram of the correlated Hubbard layer towards the overdoped regime. In other words, the diminished SC of 2D Hubbard model in the overdoped regime can be enhanced by hybridizing with an additional ``dry'' conduction band.

\section{Model and Methodology}
\subsection{Two-orbital $d$-$s$ model}
We employ the same minimal two-orbital $d$-$s$ model as our previous work~\cite{MJ2022} but our purpose here is to explore a general model incorporating two hybridized orbitals with one of them has tiny charge carrier density and the other one the conventional Hubbard band. The Hamiltonian reads as follows
\begin{align} \label{eq:HM}
	\hat{H} = & \sum_{k\sigma} (E^d_k n^d_{k\sigma} +E^s_k n^s_{k\sigma}) + U \sum_i n^d_{i\uparrow}n^d_{i\downarrow}\nonumber\\ 
	& + \sum_{k\sigma} V_k (d^\dagger_{k\sigma}s^{\phantom\dagger}_{k\sigma}+h.c.) 
\end{align}
with two orbitals' dispersion and $d$-$s$ hybridization as
\begin{align} \label{dis}
	E^d_k = & -2 t_d (\cos k_x+\cos k_y)+4 t'_d \cos k_x\cos k_y -\mu \nonumber\\ 
	E^s_k = & -2 t_s (\cos k_x+\cos k_y) + \epsilon_s -\mu \nonumber\\ 
	V_k = & -2 V (\cos k_x+\cos k_y)
\end{align}
where $d^{\dagger}_{k\sigma}(s^{\dagger}_{k\sigma})$ are electronic creation operators in momentum space for two orbitals. $n^{d(s)}_{i\sigma}$ and $n^{d(s)}_{k\sigma}$ are the associated number operators in the real and momentum spaces separately. The chemical potential $\mu$ controls the total electron density while the site energy $\epsilon_s$ of $s$ orbital (the site energy of $d$ orbital is fixed to be zero) is a tunable parameter for desired relative electron density between two orbitals. Note that we will focus on the situation of small $n_s \sim 0.1$ (``dry'' conduction band). In fact, as discussed later, all the observations reported in this work only appear in this ``dry'' regime.
The nearest-neighbor hopping between $d$ orbitals $t_d=1$ is set as the energy unit as the usual convention. Besides, typical $U/t_d=7, t'_d=-0.15$ are adopted for direct comparison with previous studies of conventional 2D Hubbard model~\cite{Maier2016,Maier2019}.

The $d$-$s$ hybridization $V_k$ is decisive for various physical properties. Without loss of generality, we only illustrate the results in the case of nearest-neighbor inter-orbital hybridization as shown in Eq.~\ref{dis}~\cite{dft25}. The case of onsite inter-orbital hybridization leads to qualitatively similar observations.  
In IL-nickelates, the hybridization between Ni and rare-earth ions is widely believed to be small~\cite{TPD_ds} while a recent investigation indicated that the dominant hybridization between Ni-3d orbitals and the effective interstitial s orbital can induce a large inter-cell hopping depending on the number of orbitals incorporated in Wannier fitting procedure~\cite{dft25}. Moreover, the recent promising experimental observation~\cite{pressure} of pressure induced enhancement of $T_c$ has direct implication to explore the regime of relatively large $V$ mimicking the pressure effects. Therefore, the present work will generically study a range of $V$ to explore its influence even though its magnitude can be unphysical for IL-nickelates. The aim is to examine the detailed interplay of various factors such as $V, n_d, n_s, t_s$ and temperature scale on the $d$-wave SC physics of $d$ orbital. Unless specified, all the major results presented here are for $n_s=0.1, t_s=0.5$ and the effects of $n_s$ and $s$ band structure will be discussed later.

\subsection{Dynamical cluster quantum Monte Carlo}
We employ dynamical cluster approximation (DCA)~\cite{Hettler98,Maier05,code} with the continuous-time auxilary-field (CT-AUX) quantum Monte Carlo (QMC) cluster solver~\cite{GullCTAUX} to
numerically solve the $d$-$s$ model Eq.~\eqref{eq:HM}.
As one of many-body quantum embedding methods, DCA calculates the physical quantities in the thermodynamic limit via mapping the bulk lattice problem onto a finite cluster embedded in a mean-field in a self-consistent manner~\cite{Hettler98,Maier05}, which is specifically realized by the convergence between the cluster and coarse-grained single-particle Green's function. Although DCA and its cousins of quantum embedding methods have provided much insights on the strongly correlated electronic systems, it has noticeable limitations mainly arising from the smaller tractable cluster size than finite-size QMC simulations albeit with better minus sign problem. 

To simulate a wide range of doping levels at the strongly correlated $U/t_d=7$ with simultaneously manageable QMC sign problem, we stick on the moderate $N_c=12$ DCA cluster for sufficient resolution in momentum space to approach $T_c$ as much as possible~\cite{Maier2022} as well as the direct comparison with previous extensive study for pure Hubbard model (without $s$ orbital)~\cite{Maier2019}. Interestingly, much more mitigated sign problems are commonly observed after turning on the $d$-$s$ hybridization so that we can reach directly or close to $T_c$, which is essential for the exploration of physics near $T_c$~\cite{Maier2022}.

The SC properties can be studied via solving the Bethe-Salpeter equation (BSE) in the eigen-equation form in the particle-particle channel~\cite{Maier06,Scalapino06}
\begin{align} \label{BSE}
    -\frac{T}{N_c}\sum_{K'}
	\Gamma^{pp}(K,K')
	\bar{\chi}_0^{pp}(K')\phi_\alpha(K') =\lambda_\alpha(T) \phi_\alpha(K)
\end{align}
where $\Gamma^{pp}(K,K')$ denotes the lattice irreducible particle-particle vertex of the effective cluster problem with combining the cluster momenta $\bf K$ and Matsubara frequencies $\omega_n=(2n+1)\pi T$ as $K=(\mathbf{K}, i\omega_n)$. DCA assumes that the desired lattice two-particle irreducible vertex $\Gamma$ can be approximated by the its cluster counterpart $\Gamma_c$, namely 
\begin{align} \label{e2}
  \chi_{c\sigma\sigma'}(q,K,K') &= \chi^0_{c\sigma\sigma'}(q,K,K') + \chi^0_{c\sigma\sigma''}(q,K,K'') \nonumber \\
  & \times \Gamma_{c\sigma''\sigma'''}(q,K'',K''') \chi_{c\sigma'''\sigma'}(q,K''',K')
\end{align}
where the cluster two-particle Green's function 
\begin{align} \label{e1}
  \chi_{c\sigma\sigma'}(q,K,K') &= \int^{\beta}_0 \int^{\beta}_0 \int^{\beta}_0 \int^{\beta}_0 d\tau_1 d\tau_2 d\tau_3 d\tau_4 \nonumber \\
  & \times e^{i[(\omega_n+\nu)\tau_1 -\omega_n\tau_2 +\omega_{n'}\tau_3 -(\omega_{n'}+\nu)\tau_4]} \nonumber \\
  \times \langle \mathcal{T} & c^{\dagger}_{K+q,\sigma}(\tau_1) c^{\phantom{\dagger}}_{K\sigma}(\tau_2) c^{\dagger}_{K'\sigma'}(\tau_3) c^{\phantom{\dagger}}_{K'+q,\sigma'}(\tau_4) \rangle
\end{align}
with conventional notation $K=(\mathbf{K},i\omega_n)$, $K'=(\mathbf{K'},i\omega_{n'})$, $q=(\mathbf{q},i\nu)$ and the time-ordering operator $\mathcal{T}$ can be calculated via a particular DCA cluster solver. In this work, we are mostly interested in the even-frequency even-parity (spin singlet) $d$-wave pairing so that $q=(\mathbf{q},i\nu)=0$ is fixed~\cite{Maier06,Scalapino06}.

Meanwhile, the coarse-grained bare particle-particle susceptibility
\begin{align}\label{eq:chipp}
	\bar{\chi}^{pp}_0(K) = \frac{N_c}{N}\sum_{k'}G(K+k')G(-K-k')
\end{align}
is obtained via the dressed single-particle Green's function $G(k)\equiv G({\bf k},i\omega_n) =
[i\omega_n+\mu-\varepsilon_{\bf k}-\Sigma({\bf K},i\omega_n)]^{-1}$, where $\mathbf{k}$ belongs to the DCA patch surrounding the cluster momentum $\mathbf{K}$, whose evaluation involves the chemical potential $\mu$,   the non-interacting dispersion relation $\varepsilon_{\bf k}$, and the cluster self-energy $\Sigma({\bf K},i\omega_n)$. 


Physically, the methodology adopted here applies for the normal state pairing tendency reflected by the magnitude of the leading eigenvalue $\lambda_\alpha(T)$ in the superconducting channel with pairing symmetry $\alpha$ by lowering down the temperature gradually until $T_c$, which is extracted as the temperature scale where the leading eigenvalue $\lambda(T_c)=1$. In the mean time, the spatial, frequency, and orbital dependence of the corresponding eigenvector $\phi_\alpha({\bf K},i\omega_n)$ can be viewed as the normal state analog of the SC gap function to manifest the structure of the pairing interaction $\Gamma^{pp}$~\cite{Maier06,Scalapino06}.
It has been widely accepted that the $d$-wave pairing dominates the cuprate superconductors and closely relevant Hubbard model~\cite{Scalapino06,Qin2022}. In the presence of an additional ``dry'' conduction $s$ band, our present study found that the leading pairing symmetry still occurs in the $d$-wave channel with momentum structure $\cos K_x - \cos K_y$. 

Importantly, determining $T_c$ via $\lambda(T_c)=1$ is only applicable if the sign problem permits us to get access to $T_c$. 
Previous studies~\cite{Maier2019} have discovered that the pairfield has strong Emery–Kivelson phase fluctuations~\cite{Kivelson1995} for dopings with a PG while BCS Cooper pair fluctuations~\cite{Abrahams1966} for dopings without a PG. Precisely, $d$-wave pairing in a 2D system occurs at a Kosterlitz–Thouless (KT) transition $T_{KT}$~\cite{Maier05a} and thereby, in the presence of PG, there is a linear $T$-dependence regime of $1-\lambda_d(T)$ at moderate temperatures associated with the pairfield's phase fluctuations followed by crossing over to the lower temperature vortex–antivortex KT exponential $T$-dependence. Hence, the $T_{KT}$ can be determined by the exponential fitting of $1-\lambda_d(T)$ in underdoped regime with PG. In contrast, at higher hole doping regime without PG behavior in the pure Hubbard model, $1-\lambda_d(T)$ shows logarithmic BCS form $\ln(T/T_{MF})$ where the mean-field temperature $T_{MF}$ is close to $T_{KT}$. Therefore, when the severe sign problem prevents from approaching to low enough temperature, we rely on different fitting procedure of the temperature evolution of $1-\lambda_d(T)$.

\section{Results}
\begin{figure} 
\psfig{figure=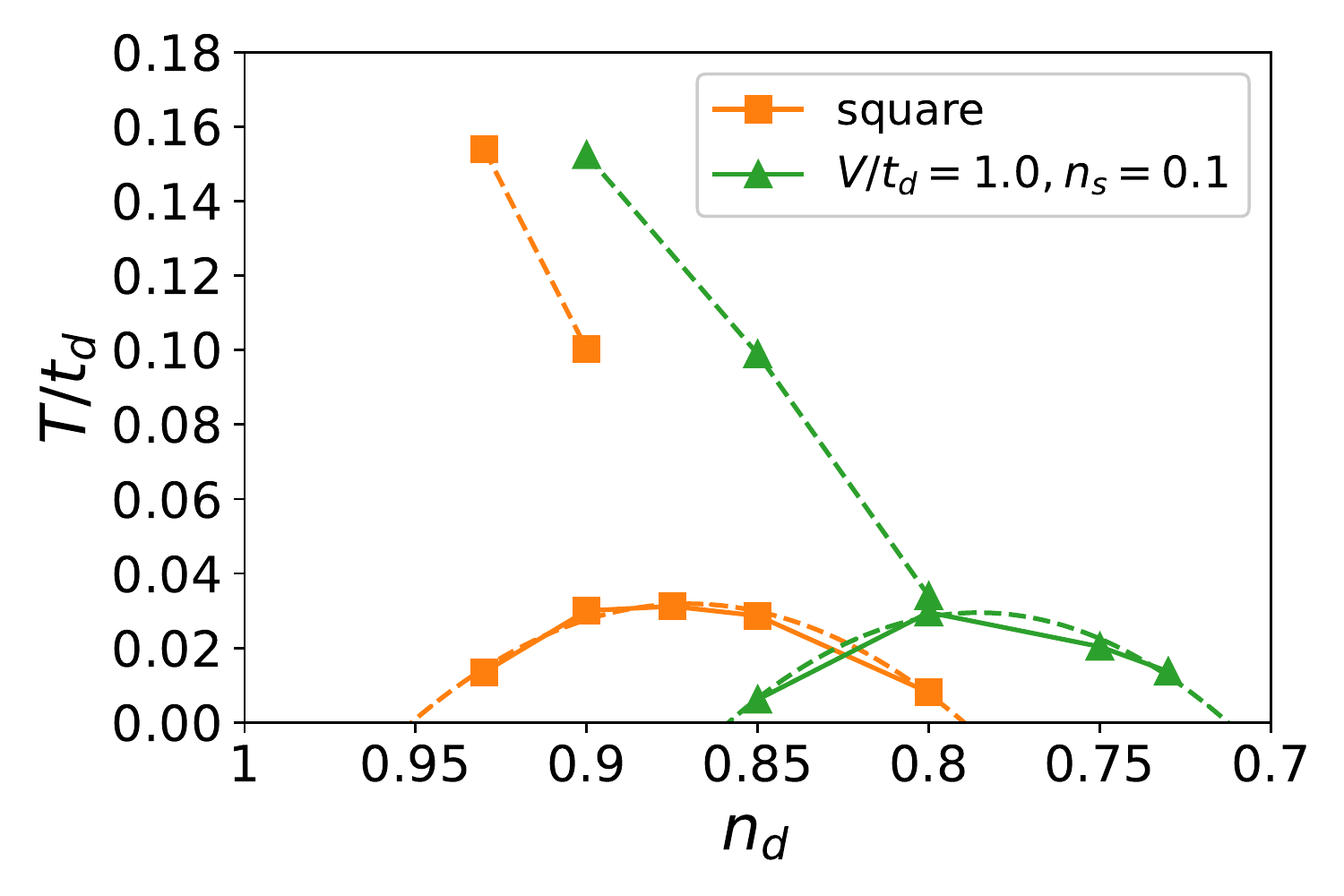,height=5.8cm,width=.49\textwidth, clip} 
\caption{Phase diagram of $d$-$s$ model illustrating the shift of the SC dome (Kosterlitz–Thouless transition line) to the overdoped regime by moderate $d$-$s$ hybridization $V/t_d=1.0$ and dry $s$ band. The $d$-wave SC can thus be enhanced (suppressed) at highly (lightly) doped regime. The pseudogap temperature scale $T^*$ shifts together with $T_{KT}$ and terminates around the optimal doping in both pure Hubbard and $d$-$s$ models. Other parameters are fixed $U/t_d=7, t'_d=-0.15, t_s=0.5, N_c=12$.}
\label{phase}
\end{figure}

Figure~\ref{phase} summarizes the phase diagram of $d$-$s$ model as our main result, where the green dome labels the $d$-wave pairing KT transition $T_{KT}$ line. The other monotonically decreasing curve denotes the PG temperature scale $T^*$. As comparison, the orange curves provide the corresponding evolution of $T_{KT}$ and $T^*$ in the conventional 2D Hubbard model.
Obviously, the seemingly negligible dry $s$ conduction band has remarkable shifting effect on the SC dome, which is reminiscent of the effects of increasing the magnitude of next nearest neighbor $t'$ in the Hubbard model~\cite{Jarrell2013}. This clearly provide a reasonable route of promoting the $d$-wave SC in the overdoped regime. 
Interestingly, the pseudogap $T^*(n_d)$ curve shifts to the higher doping accordingly, suggesting the same origin of its shift as SC dome.
More importantly, our simulations indicated that only small $n_s$ can induce these shift. As can be seen in red line, for $n_s=0.2$, the $T_{KT}$ has diminished and we confirm that even larger $n_s\sim 0.3$ completely destroys the $d$-wave pairing tendency at $d$-$s$ hybridization $V/t_d=1.0$ adopted here.  Unfortunately, the highest $T_{KT}$ in the $d$-$s$ model does not exceed the pure 2D Hubbard model, which implies that simply coupling to a dry conduction band is not a promising strategy of enhancing the SC ``globally'' in the whole doping regime.

\begin{figure*} 
\psfig{figure=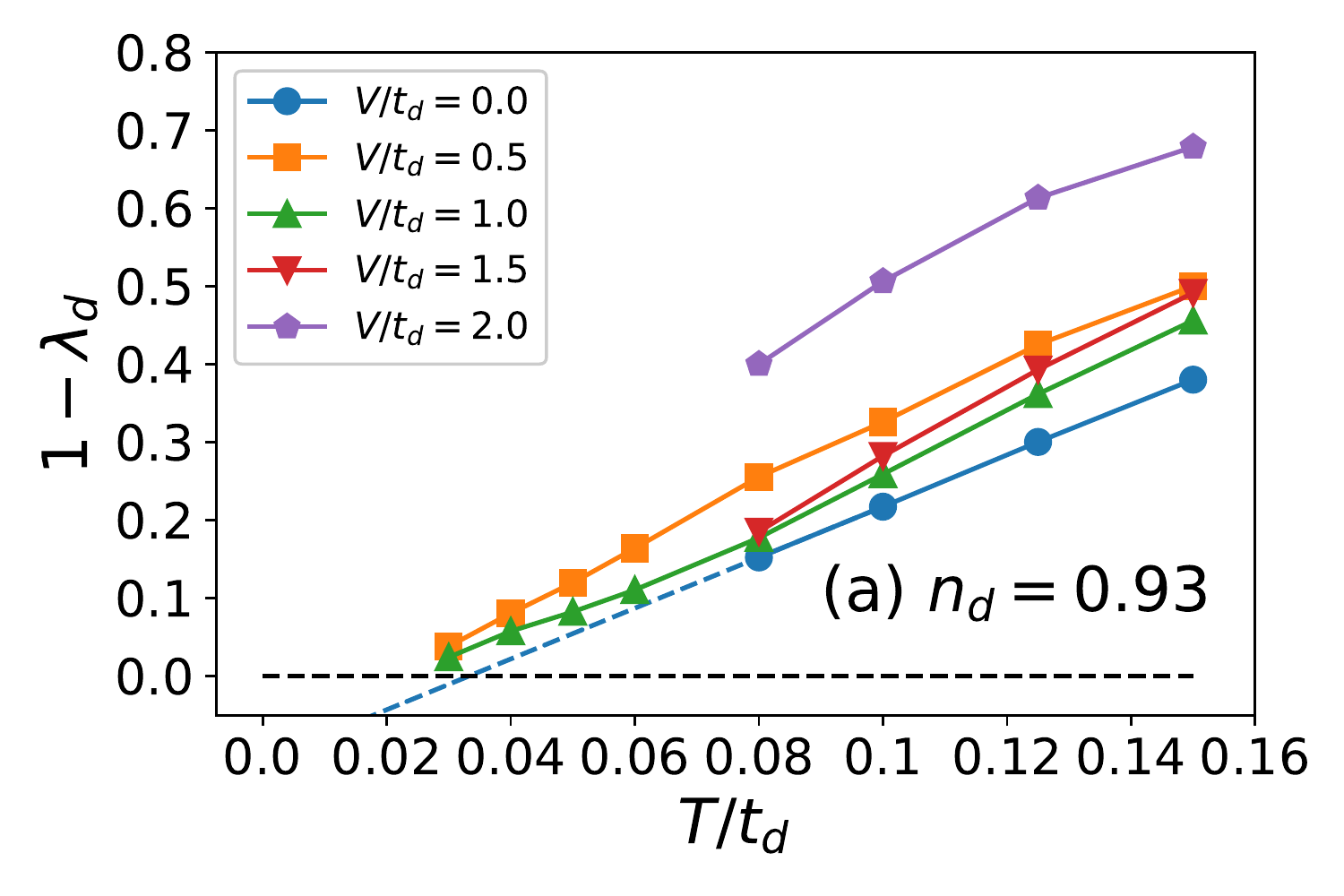,height=5.8cm,width=.49\textwidth, clip} 
\psfig{figure=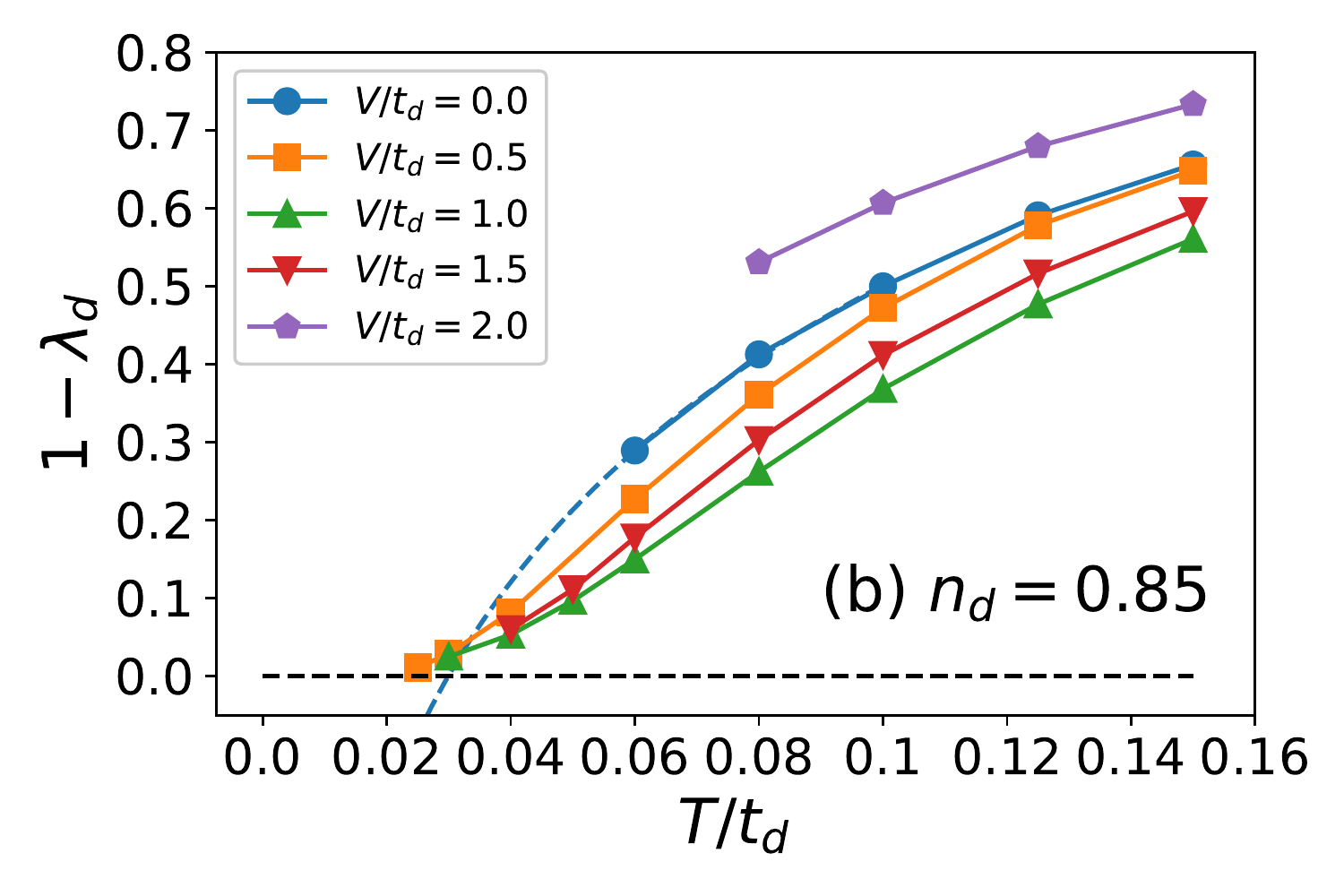,height=5.8cm,width=.49\textwidth, clip} 
\psfig{figure=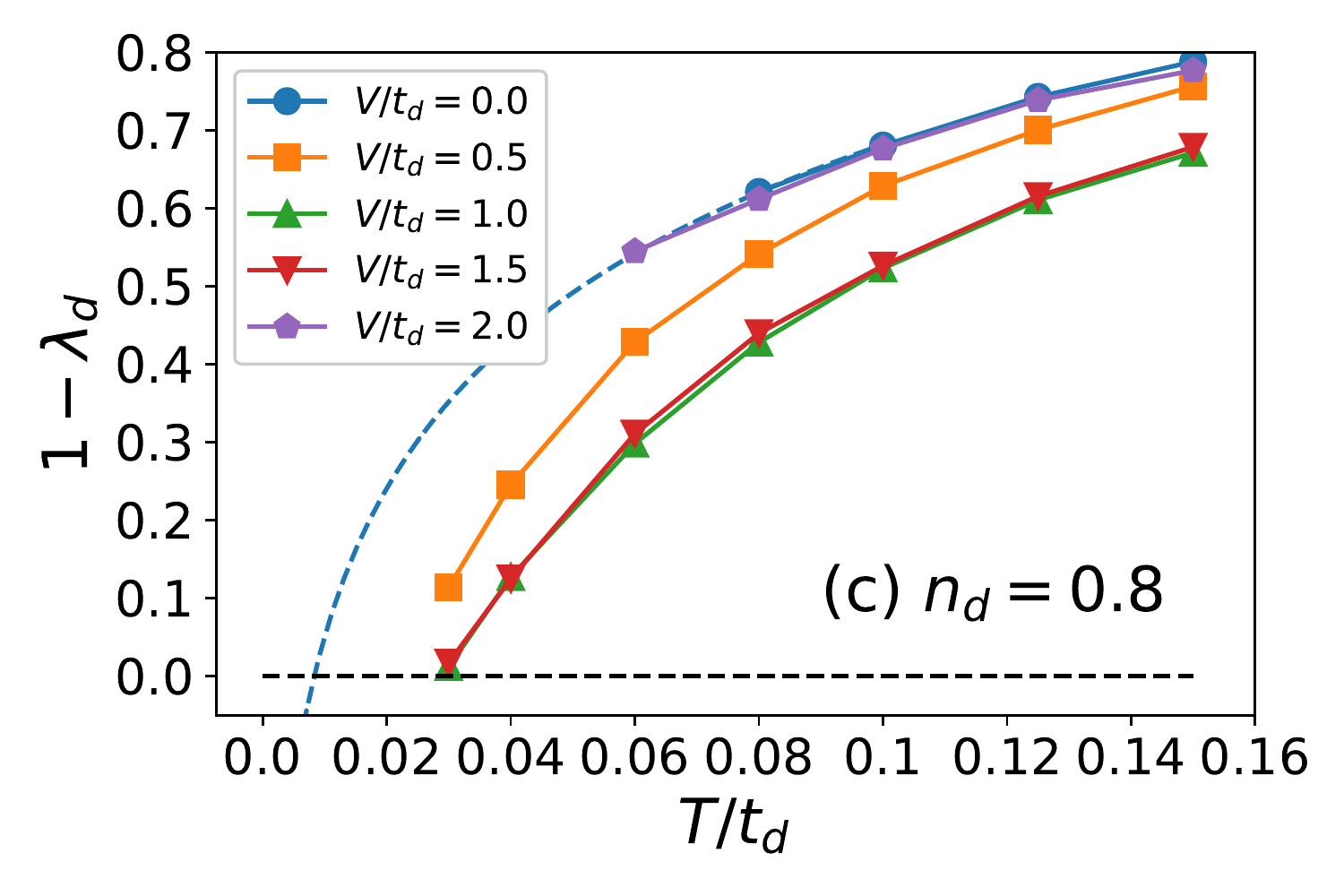,height=5.8cm,width=.49\textwidth, clip}
\psfig{figure=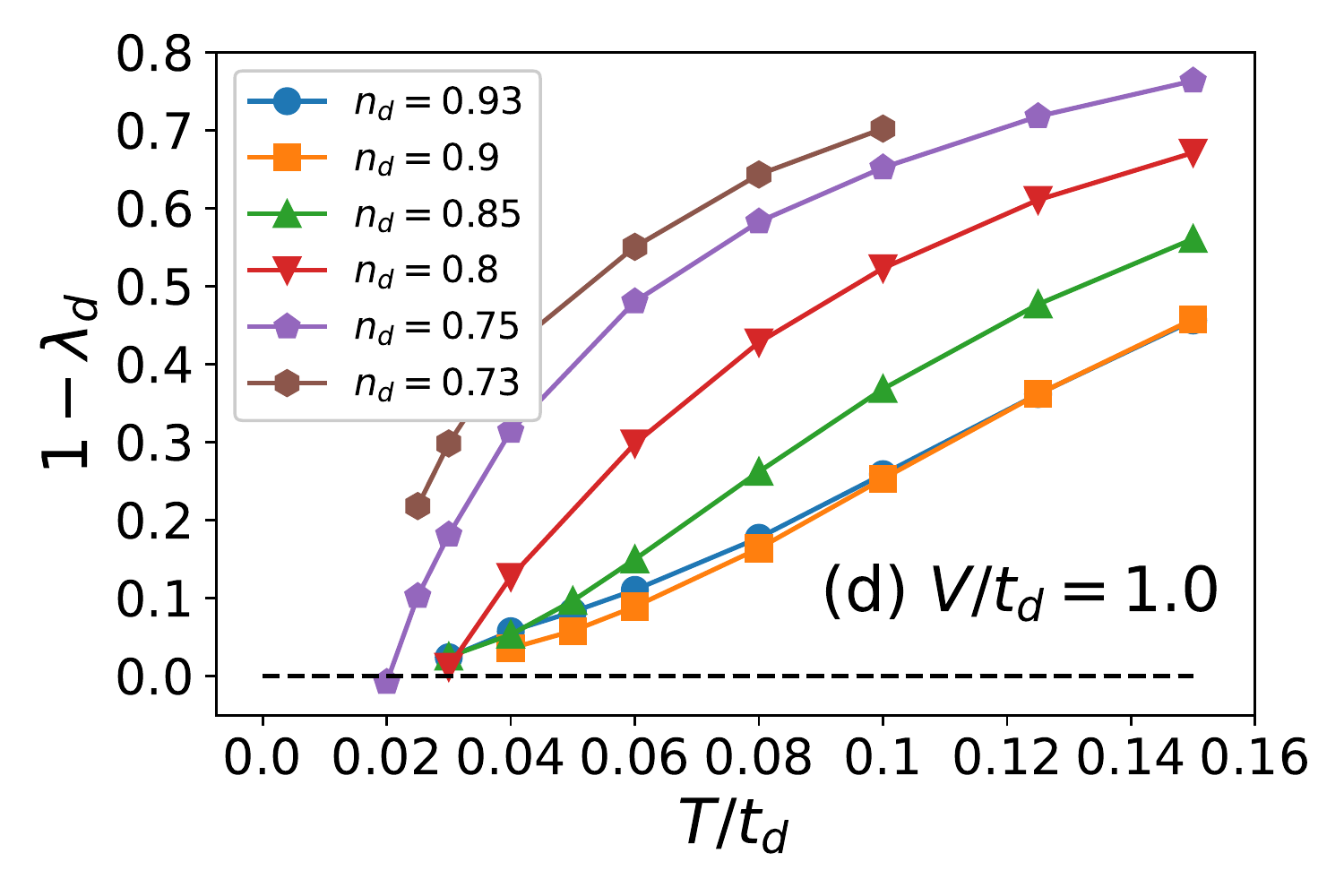,height=5.8cm,width=.49\textwidth, clip} 
\caption{Temperature evolution of $1-\lambda_d(T)$ for (a-c) three characteristic dopings reflecting distinct nature of the pair-field fluctuations associated with the presence or absence of PG. Panel (d) provides the gradual evolution at fixed hybridization $V/t_d=1.0$, which results in the KT transition line $T_{KT}(n_d)$ of Fig.~\ref{phase}. All the simulations are performed for $n_s=0.1$.}
\label{lambdaT}
\end{figure*}

In the following, we provide detailed numerical evidence for the phase diagram. Fig.~\ref{lambdaT} demonstrates the temperature evolution of $1-\lambda_d(T)$ for various doping regime. At underdoped $n_d=0.93$ (panel a), turning on $V$ suppresses the pairing tendency in spite of the mitigated trend at $V/t_d=1.0$.
The dominant feature is the change of its curvature as lowering the temperature compared with the pure Hubbard model ($V=0$). As discussed earlier, this indicates strong Emery–Kivelson phase fluctuations~\cite{Kivelson1995} in the presence of PG. Hence, the $T_{KT}$, for example at $V/t_d=1.0$, can be determined by the exponential fitting of $1-\lambda_d(T)$. In fact, the extrapolated $T_{KT}$ is extermely low, if not zero. Unfortunately, larger $V$ prevents us from accessing to lower temperature owing to the severe QMC sign problem.

At higher hole doping $n_d=0.85$ (panel b), namely without PG behavior in the pure Hubbard model, $1-\lambda_d(T)$ shows logarithmic BCS form $\ln(T/T_{MF})$ where the mean-field temperature $T_{MF}$ is close to so that can approximate $T_{KT}$. 
Turning on moderate $V$ induces stronger pairing tendency as evidenced by the curves below the blue one. 
Nonetheless, the naive expectation of higher $T_{MF}$ turns out to be false at low enough temperature, where the curvature qualitatively changes from the BCS logarithmic form at $V=0$ to the linear followed by an exponential KT behavior normally observed for the system with PG~\cite{Maier2019}. In fact, as shown in the phase diagram Fig.~\ref{phase}, $n_d=0.85$ in $d$-$s$ model corresponds to the doping regime with PG. In other words, the additional dry $s$ band strengthens the pairfield's phase fluctuation of $d$-orbital so that the BCS type Cooper pair fluctuation transforms to Emery-Kivelson phase fluctuation.
As a byproduct, this provides a cautious lesson for various many-body computational techniques on analyzing any indicators of the pairing tendency such as $\lambda_d(T)$ or pairfield susceptibility, whose behavior at relatively high temperatures can be misleading for drawing conclusion on the physical pairing instability.

As one further enters into even higher hole doping regime, e.g. $n_d=0.8$ (panel c), all curves exhibit BCS logarithmic form, suggesting that the highly doped regime is dominated by the BCS Cooper pair fluctuations regardless of the existence of dry $s$ band. Noticeably, the enhancement of the $d$-wave pairing is strongest in this overdoped regime, as indicated by the green and red curves for moderate $V/t_d=1.0-1.5$.

Summarizing the above three characteristic doping cases, the hybridization $V$ has non-monotonic influence on the pairing tendency, which is the most efficiently promoted by moderate $V$. Undoubtedly, large enough hybridization $V$ will completely destroy or largely weaken the pairing tendency as indicated by the curves for $V/t_d=2.0$. As discussed later, since the $d$-band's charge carriers are holes while the additional dry $s$ band lives electrons, it can be reasonably believed that the ``light'' electron-``heavy'' hole binding plays an important role, which is reminiscent of the conventional Kondo screening relevant for heavy fermion systems. 

Now that the optimal hybridization for enhancing the pairing tendency occurs around ``magic'' $V/t_d=1.0$, panel (d) shows the temperature dependence of $1-\lambda_d(T)$ for various hole dopings at fixed $V/t_d=1.0$. Obviously, increasing the hole doping drastically modifies the curvature at low temperature associated with the transition from pairfield phase fluctuation to BCS Cooper pair fluctuation so that the corresponding $T_{KT}(n_d)$ supports a dome as shown in the phase diagram Fig.~\ref{phase}.

\begin{figure} 
\psfig{figure=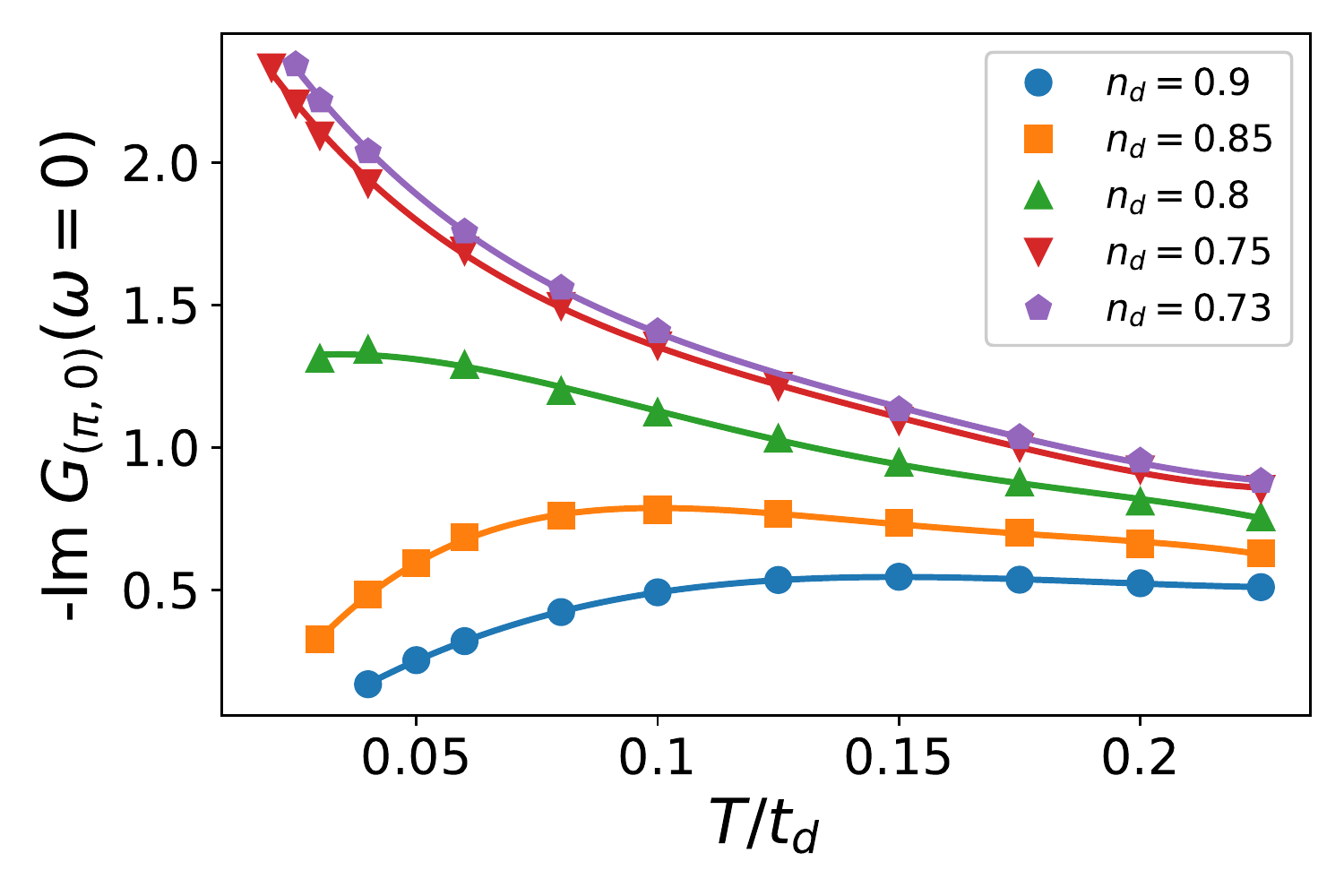,height=5.8cm,width=.49\textwidth, clip}
\caption{$T$-dependence of extrapolated antinodal zero-frequency spectral function obtained from a linear extrapolation to zero frequency of -Im$G_{(\pi,0)}(i\omega_n)$ at the first two Matsubara. Its decrease at low temperature signifies the presence of PG phenomena matching with the phase diagram Fig.~\ref{phase}. $V/t_d=1.0, n_s=0.1$ are fixed.}
\label{PG}
\end{figure}

We have mentioned that the curvature of $1-\lambda_d(T)$ reflects the presence or absence of PG feature. As a well-known puzzle, PG phenomenology within the underdoped regime of the phase diagram of cuprates is believed to profoundly affect SC phase so that PG is even more puzzling than SC itself in some sense. Hereby, we continue to explore the PG behavior in $d$-$s$ model, especially the influence of hybridization $V$ on the PG temperature scale. To determine the PG $T^*$, we rely on the temperature scale where the antinodal zero-frequency spectral function starts to decrease with $T$, which is in turn obtained from a linear extrapolation of -Im$G_{(\pi,0)}(i\omega_n)$ at the first two Matsubara frequencies to zero frequency~\cite{WeiWu2022}. Fig.~\ref{PG} demonstrates the $T$-dependence of extrapolated antinodal extrapolated -Im$G_{(\pi,0)}(0)$, whose decrease at low temperature signifies the opening of a spectral gap in the antinodal direction. Evidently, the PG in the lightly doped regime gradually disappears with increasing the doping level, which is quite similar to the trend observed in the pure Hubbard model~\cite{Maier2019}. The only difference lies in the crossover doping, which shifts to higher doping $\sim 0.8$ in the presence of dry $s$ band compared to $\sim 0.9$ of pure Hubbard model. Accordingly, the PG line $T^*(n_d)$ in Fig.~\ref{phase} shows the right shift towards the highly doping levels accompanying with the movement of $T_{KT}(n_d)$ dome. 

\section{Discussion}
\subsection{Effective pairing interaction and pair-field susceptibility}
With the above numerical observation, we next examine their implication and possible physical origin. The dominant feature to be analyzed is the non-monotonic dependence of the $d$-wave pairing tendency upon both the hybridization magnitude $V$ and doping level $n_d$.

\begin{figure}[t]
\psfig{figure=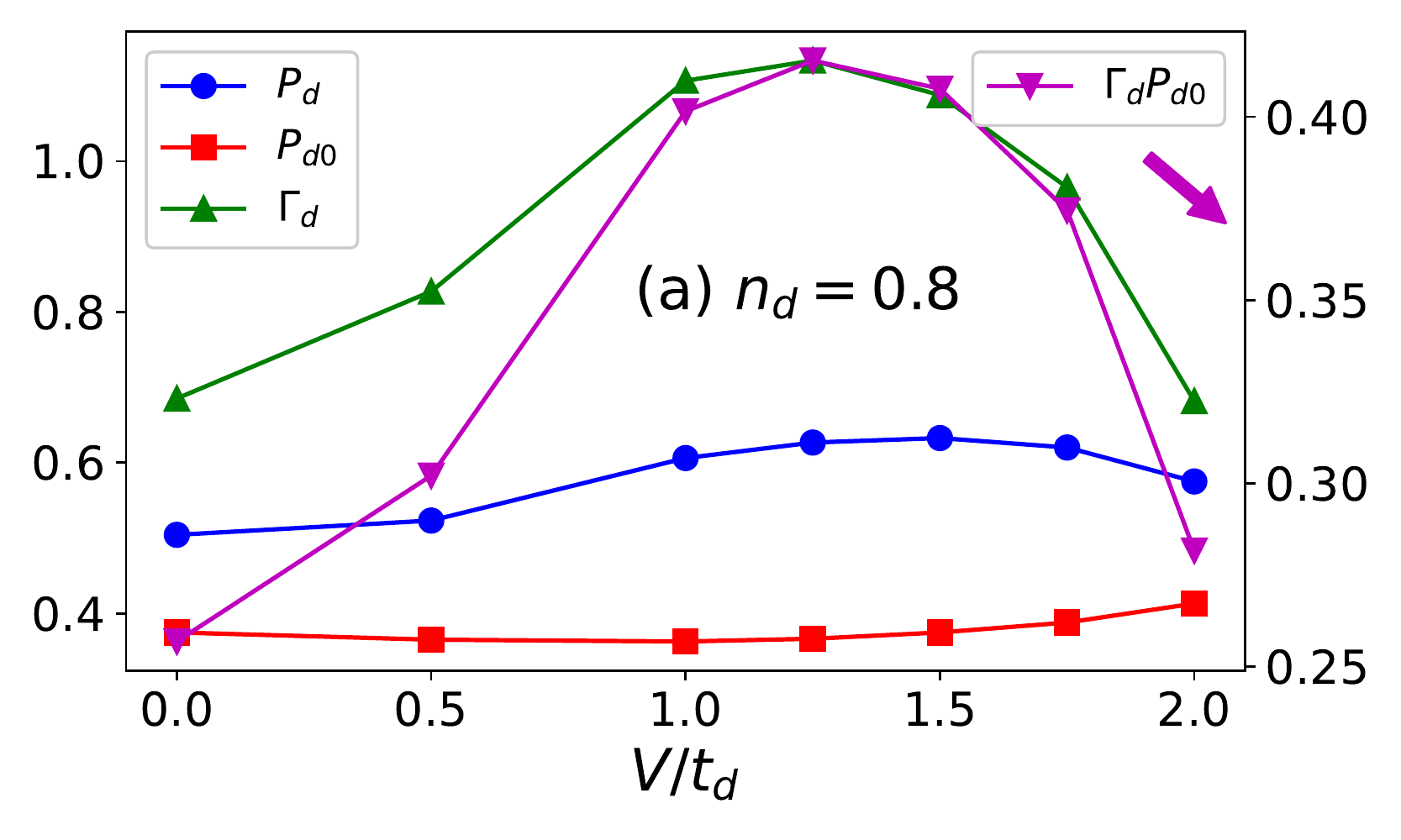,height=5.5cm,width=.49\textwidth, clip}
\psfig{figure=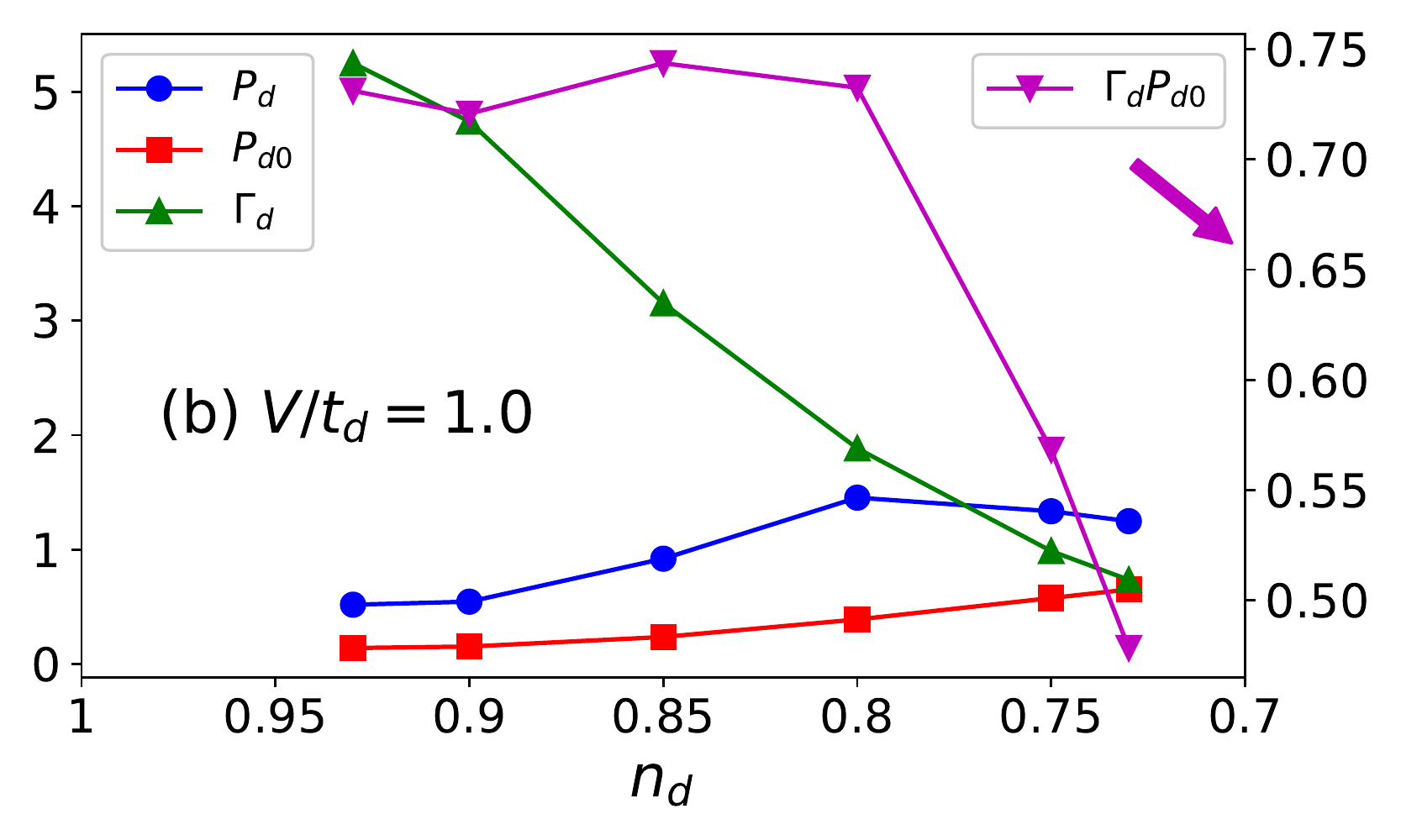,height=5.5cm,width=.49\textwidth, clip} 
\caption{Effective $d$-wave pairing interaction, bare and dressed pair-field susceptibility versus (a) $V$ for fixed $n_d=0.8$ at $T/t_d=0.08$ and (b) $n_d$ at fixed $V/t_d=1.0$ at $T/t_d=0.04$.}
\label{Pd}
\end{figure}

Physically, the driver of pairing in a particular symmetry channel can be described by the effective pairing interaction and the intrinsic pair-field susceptibility projected onto that particular symmetry. Therefore, we calculated the $d$-wave projection of pair-field susceptibility $P_d$ following the standard procedure developed earlier~\cite{Jarrell01,Jarrell2013}. The corresponding bare pair-field susceptibility $P_{d0}$ can also be easily obtained as the $d$-wave projection of product of fully dressed
single-particle Green functions similar to $\bar{\chi}_0^{pp}$ in Eq.~\ref{BSE}. With these two quantities, an effective $d$-wave pairing interaction $\Gamma_d$ can be defined as
$\Gamma_d(T) = P^{-1}_{d0}(T)-P^{-1}_{d}(T)$. This analysis is essentially similar to the BSE discussed above but with the major difference that Eq.~\ref{BSE} includes all symmetry channels instead of sole $d$-wave pairing. Putting in another way, $\Gamma_d$ and $P_{d0}$ are two decomposed factors with similar roles as $\Gamma^{pp}$ and $\bar{\chi}_0^{pp}$ in BSE Eq.~\ref{BSE} respectively.
Thus, it is not surprising that the product $\Gamma_dP_{d0}$'s dependence on $V$ and $n_d$ will qualitatively track the behavior of $\lambda_d$ obtained by BSE. This is confirmed by Fig.~\ref{Pd}, where the purple lines in two panels shows its non-monotonic evolution with $V$ and $n_d$ separately, which exactly follows the nonmonotonicity of $\lambda_d$ in Fig.~\ref{lambdaT}. Note that the optimal $T_{KT}$ occurs at $n_d=0.8$ in Fig.~\ref{phase} because the analysis of Fig.~\ref{Pd} is performed at the lowest available $T$ scale while $T_{KT}$ is extrapolated via distinct pair-field fluctuations upon the presence or absence of PG. 

Moreover, Fig.~\ref{Pd} illustrates the behavior of $P_d$, $P_{d0}$, and $\Gamma_d$ separately versus (a) the hybridization for fixed $n_d=0.8$ and (b) $n_d$ at fixed $V/t_d=1.0$, which reveals some distinct feature between these two types of non-monotonicity.
For fixed $n_d=0.8$ (panel a), although $P_d$ generally shows similar broad bump as $\Gamma_dP_{d0}$, the bare $P_{d0}$ is almost independent on $V$ until quite large magnitude. This results in the effective interaction $\Gamma_d$ tracking $P_d$'s nonmonotonicity. Therefore, the origin of the strongest enhancement of $d$-wave pairing at moderate hybridization $V$ directly lies in the strongest pairing interaction $\Gamma_d$ (green curve).
Conversely, for fixed moderate $V/t_d=1.0$ (panel b), the SC dome originates from the compromise between monotonically decreasing pairing interaction $\Gamma_d$ and the gradual increase of the bare pair-field susceptibility $P_{d0}$, whose physics is similar to the recent report of hybridization enhanced $s$-wave pairing with attractive Hubbard interaction~\cite{Maier2022}.

In fact, we have checked that the compromise between the opposite monotonic dependence of  $\Gamma_d$ and $P_{d0}$ also occurs in the pure Hubbard model (not shown here), which implies that, in terms of the pairing interaction, the basic physics of doping induced SC dome is regardless of the presence or not of the dry $s$ band.
On the contrary, the non-monotonicity associated with the distinct influence of the hybridization $V$ from the dry $s$ band at overdoped regime $n_d=0.8$ (panel a) mimics the situation of conventional PAM, where there exists optimal hybridization $V$ to induce AF correlation between local moments via RKKY mechanism~\cite{Mi20}.

\subsection{BSE eigenfunction and spin fluctuation}
\begin{figure*}[t]
\psfig{figure=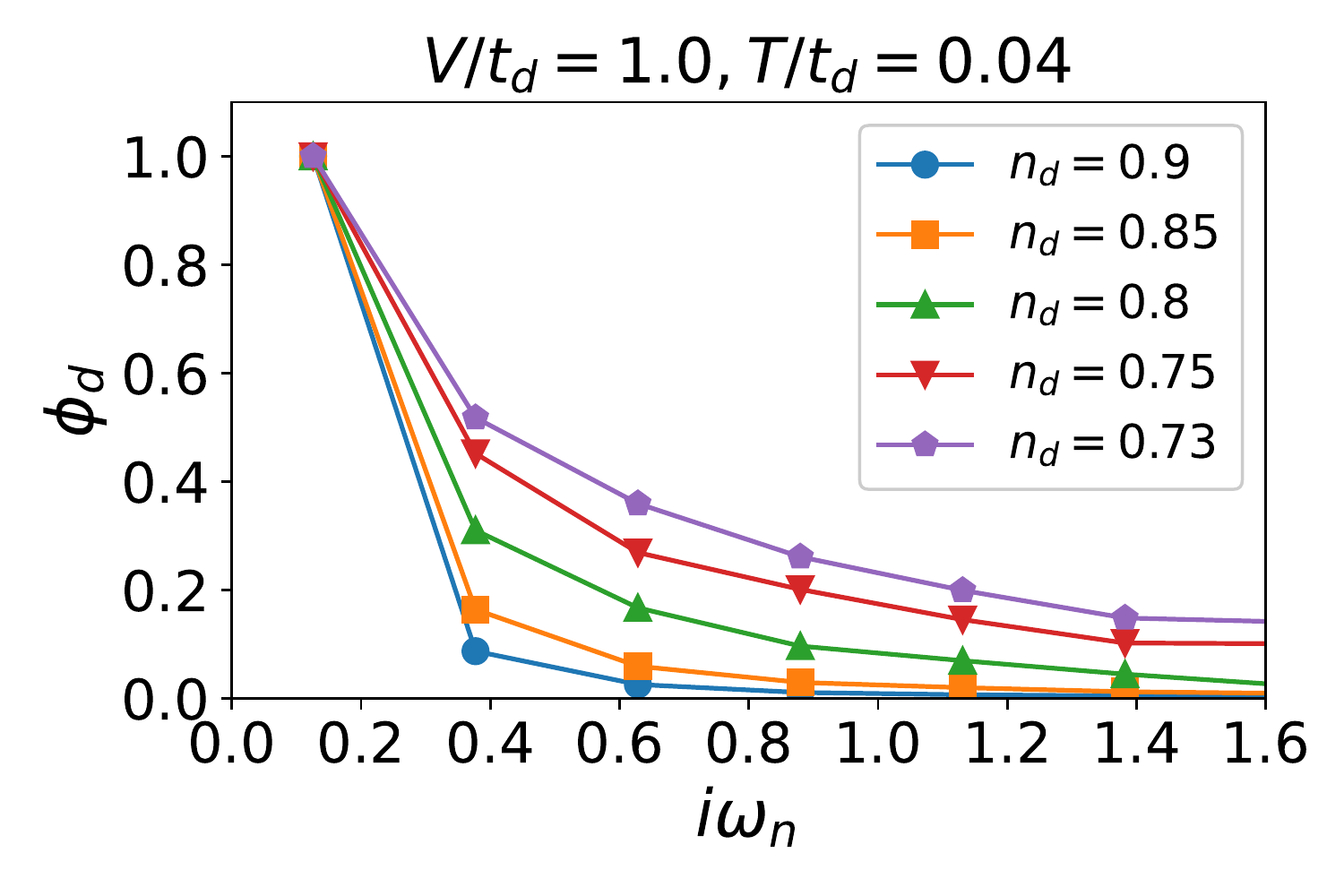,height=5.9cm,width=.49\textwidth, clip}
\psfig{figure=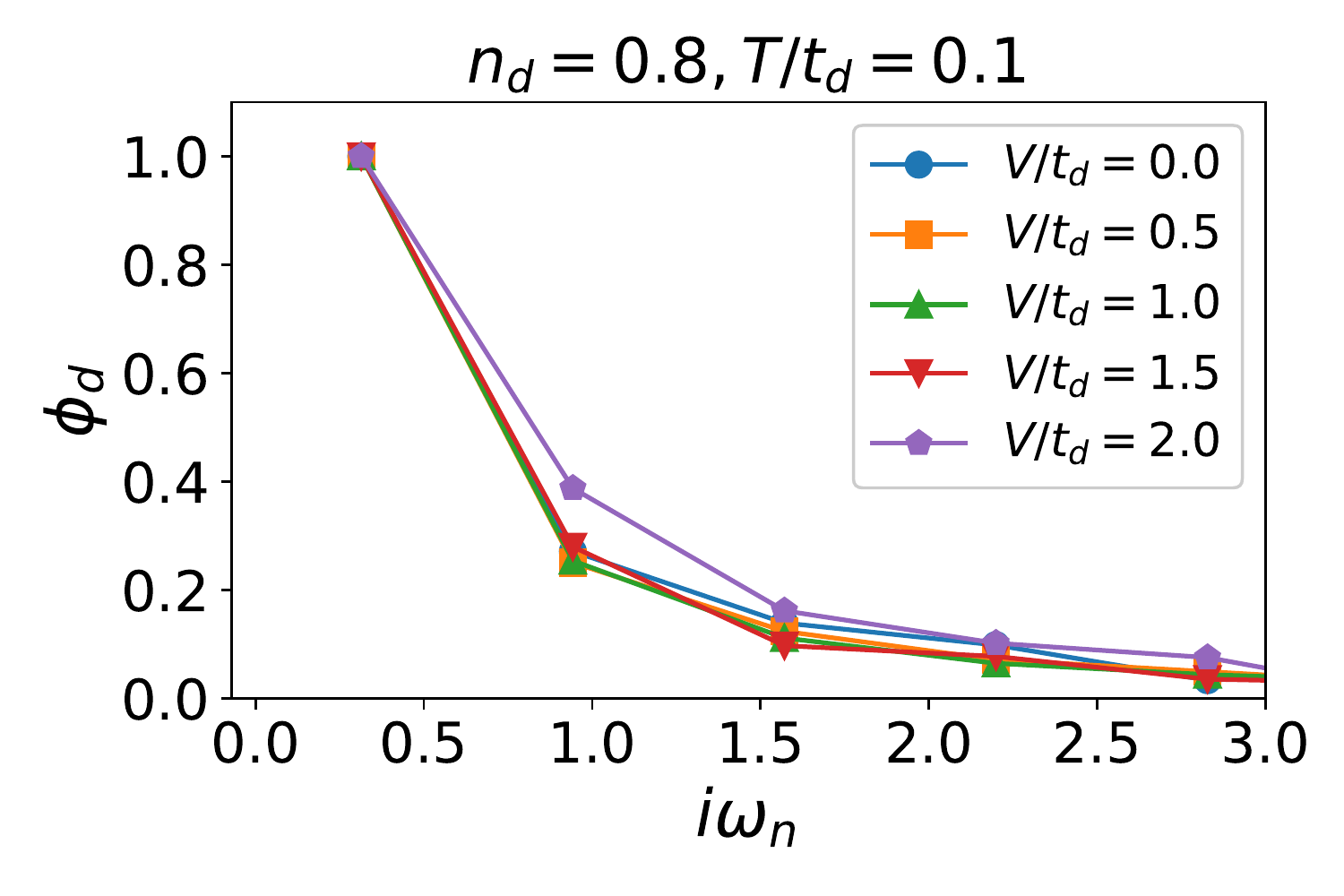,height=5.9cm,width=.49\textwidth, clip}
\psfig{figure=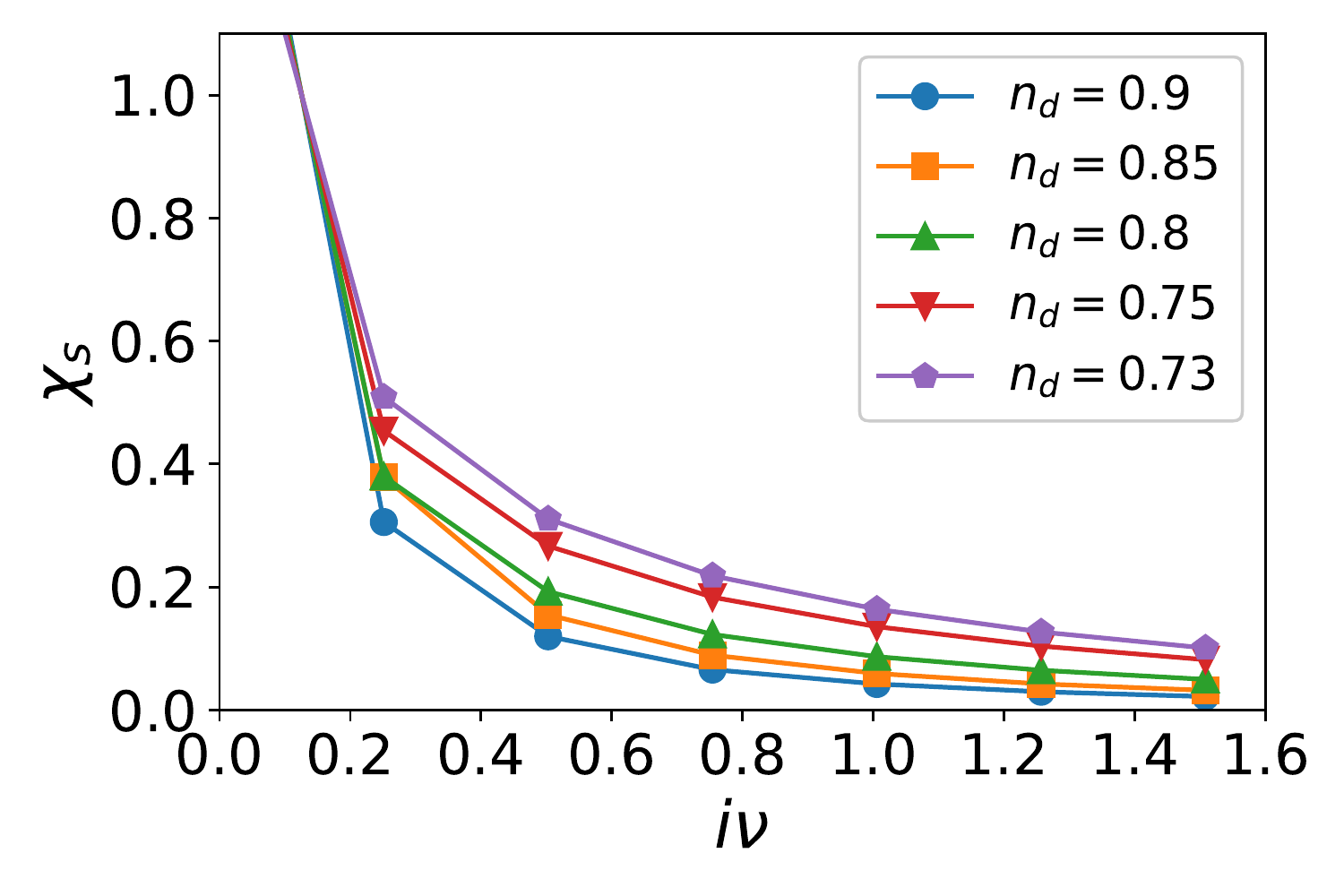,height=5.5cm,width=.49\textwidth, clip}
\psfig{figure=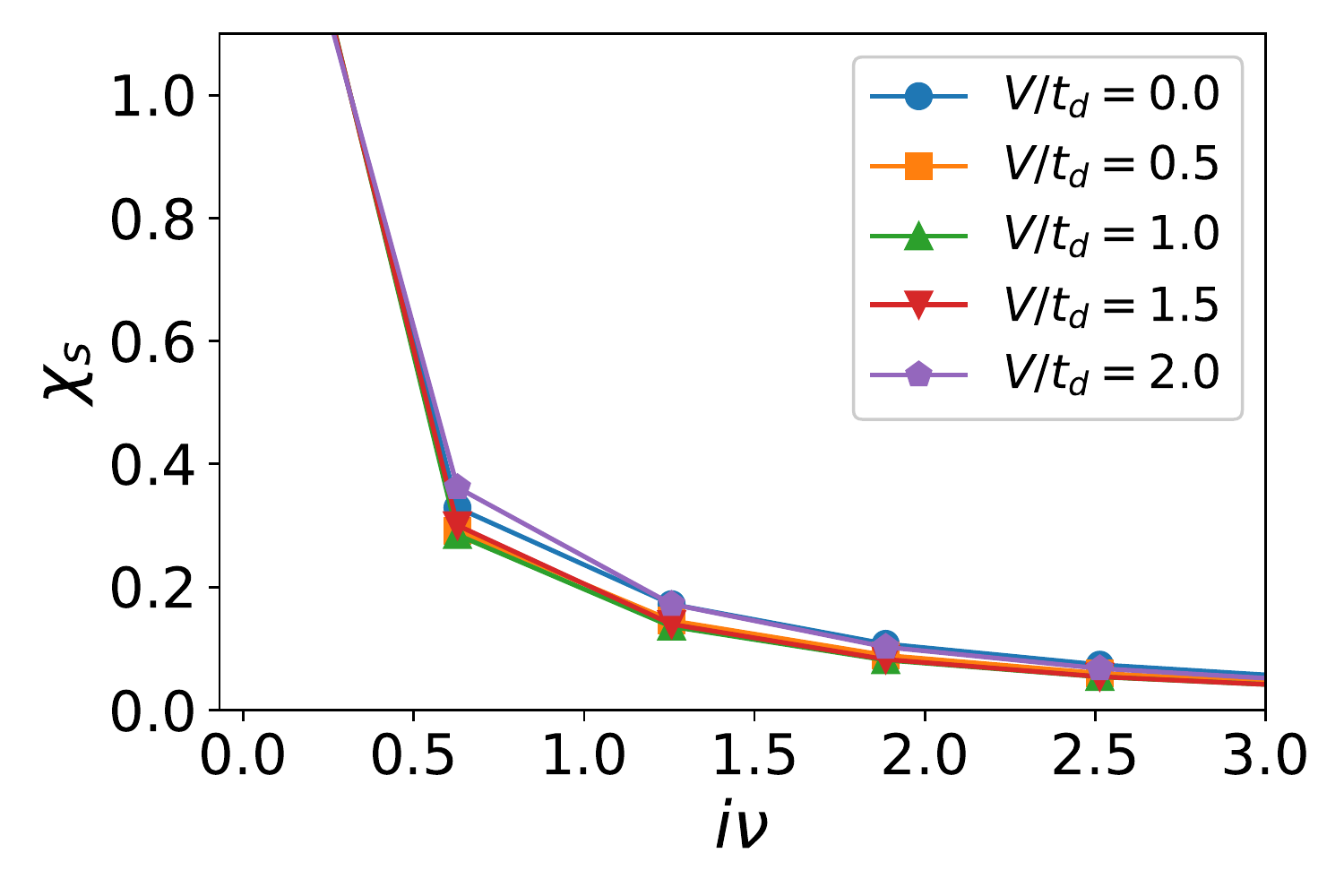,height=5.5cm,width=.49\textwidth, clip}
\caption{Frequency dependence of the renormalized leading $d$-wave BSE eigenfunction $\phi_d(i\omega_n)/\phi_d(i\pi T)$ at $\mathbf{K}=(\pi,0)$ and the AF spin susceptibility $2\chi_s(i\nu)/[\chi_s(0)+\chi_s(i\pi T)]$  at $\mathbf{q}=(\pi,\pi)$ versus (Left) $n_d$ at fixed $V/t_d=1.0$ and (Right) versus $V$ at fixed $n_d=0.8$.}
\label{chi}
\end{figure*}

Until now, we mostly gathered information from the BSE eigenvalues $\lambda_d$ in the $d$-wave channel. As mentioned before, the BSE eigenfunction provides more underlying physics as the normal state analog of SC gap function, whose momentum and frequency dependency compared with the spin susceptibility provided strong numerical evidence of the antiferromagnetic (AF) spin fluctuation's role in mediating the Cooper pairing~\cite{Maier06,Scalapino06,ScalapinoRMP}.
Furthermore, there is strong evidence that the origin of pseudogap physics of cuprates is also closely related to the AF spin fluctuation~\cite{ScalapinoRMP,Macridin06,Gull2010,Toschi2015}, although its precise role deserves further exploration~\cite{MillisGull2022,Mona2014}.

Fig.~\ref{chi} compares the frequency dependence of the renormalized leading $d$-wave BSE eigenfunction $\phi_d(\mathbf{K}=(\pi,0))$ and the AF spin susceptibility $\chi_s(\mathbf{q}=(\pi,\pi))$, which can be calculated in the particle-hole channel in the similar manner as pair-field susceptibility discussed earlier~\cite{Jarrell01}, corresponding to the two cases of Fig.~\ref{Pd}. As shown in the left panels, at fixed $V/t_d=1.0$ and a temperature scale very close to $T_{KT}$, both the eigenfunction $\phi_d(i\omega_n)$ and spin $\chi_s(i\nu)$ show monotonic evolution on $n_d$, which reflects the monotonic decrease of effective pairing interaction $\Gamma_d$ with increasing hole doping in Fig.~\ref{Pd}(b). Given that the optimal $T_{KT}$ occurs at $n_d=0.8$ shown in Fig~\ref{phase}, this implies that the retardation associated with the decay rate of $\phi_d$ and $\chi_s$ with frequency has to be at an intermediate scale for optimal SC.  
Besides, this clearly indicates the role of spin-fluctuation in mediating the $d$-wave pairing reminiscent of the pure Hubbard model~\cite{Maier06}. 
Unfortunately, our present analysis could not directly evaluate and decipher the anomalous self-energies to examine the precise role of spin-fluctuation in accounting for SC~\cite{MillisGull2022}. 

Now switching to the right panels, limited by the severe sign problem mainly in pure Hubbard model, we are restricted to relatively high temperature $T/t_d=0.1$, where both $\phi_d$ and $\chi_s$ demonstrate non-monotonic dependence on $V$, which matches well with the peak feature of $\Gamma_d$ shown in Fig.~\ref{Pd}(a). In particular, the optimal $V/t_d\sim 1.0$ induces their fastest decay, which is connected with the reduction of the characteristic frequency scale. Interestingly, combining all situations in Fig.~\ref{chi}, it can be noticed that the characteristic frequency scale seems to be inversely proportional to the pairing interaction $\Gamma_d$ in Fig.~\ref{Pd}, namely larger $\Gamma_d$ corresponds to smaller frequency scale (faster decay). Note that in the strong coupling picture, this characteristic frequency scale is set by the effective Heisenberg exchange $J \sim 4t^2/U$ in the pure Hubbard model~\cite{Scalapino06}. In the presence of $d$-$s$ hybridization $V$, the exact relation of this scale to $U$ and $V$ is complicated and deserves further examination. 
In terms of the optimal $V/t_d\sim 1.0$, we remark that at moderate hybridization, the aforementioned RKKY interaction in conventional PAM might still apply by promoting the AF correlation between $d$-orbital's local moment even when the $s$ band is quite dry so that the pairing could be possibly in turn enhanced by these AF correlations.

\subsection{``Dryness'' and band structure of $s$ band}
One important feature manifested in the phase diagram Fig.~\ref{phase} is that only ``dry'' $s$ band has all the above discussed unusual influence on the $d$-wave pairing of $d$-orbital. In fact, we have checked that larger $n_s$ would suppress the pairing. 
It is also found that larger $t_s=1.0$ so that larger $s$ bandwidth turns out to be insignificant to even quantitatively modify the results for $t_s=0.5$ presented here, which is probably related to the dryness of the $s$ band. In addition, the change from the current nearest-neighbor hybridization $V$ to onsite hybridization would not qualitatively modify our results but the optimal hybridization would be larger, namely stronger $d$-$s$ hybridization is required to efficiently promote SC.
One more interesting observation is that if we switch the sign of $t_s$ so that the dry $s$ band becomes hole-like, there is no SC enhancement in the same parameter range reported here.

Based on all these observation, one naive speculation is that the small amount of additional ``light'' $s$ electrons would bind with the doped ``heavy'' $d$ holes, which is analogous to the conventional Kondo screening relevant for heavy fermion systems, so that the effective number of mobile holes responsible for pairing decreases and thereby the parallel shifted SC dome. If this holds true, it would be expected that larger $n_s$ would require more holes to be bound together, for instance $n_d=0.6, n_s=0.3$, to reduce the effective hole doping for shifting the SC dome. Nonetheless, our simulations do not confirm this expectation. We postulate that the electron-hole binding induced AF correlation (via RKKY mechanism) between local $d$ moments transforms from short-range that can promote SC at small $n_s$ to longer-range with too much $s$ electrons, which competes with so that gradually exceeds the pairing tendency.
Naturally, too strong electron-hole binding acts as strong scatterers to weaken the pairing, akin to the Kondo singlet formation induced insulating behavior. More analysis along this line of thoughts is worthwhile, for example the possibility of the exciton formation and its relevance to our observation.

\subsection{Relevance to realistic systems}
Because our original motivation of exploring the role of a dry $s$ band came from the IL-nickelates, it is valuable elaborating on some thoughts here.  
Although there have been proposals that in IL-nickelates the hole doping would push the additional tiny electronic band mainly living in rare-earth layer to far above Fermi energy so that the system could be effectively described by an one-band picture~\cite{oneband,dft101,dft4}, our investigation demonstrated that the orbitals (currently simplified by $s$-orbital) in rare-earth layer can play a significant role in modifying the SC properties of NiO$_2$ layer by a sizeable hybridization~\cite{MMG2022,Hanghuireview}. This can be treated as mimicking the recent high pressure experiment~\cite{pressure}, where the pressure induced monotonic enhancement of $T_c$ probably originates from that the shrunk c-axis lattice constant by pressure will enhance the hybridization between the Ni-3d and rare-earth's 5d orbitals. 

Although our discovery of non-monotonic dependence of $T_{KT}$ upon hybridization $V$ contradicts with the experimentally monotonic enhancement of $T_c$, our simulations indicate that tiny amount of $s$ electrons can indeed have drastic effects of shifting SC dome, which seemingly matches the trend from cuprates to IL-nickelates~\cite{2019Nature}, where the optimal hole doping occurs at a much higher level than cuprates. Despite that the dry $s$ band cannot promote the highest $T_c$, this provides a new route of enhancing SC especially in the overdoped regime. More generally, the presence of the additional bands may hint on a new way of tuning SC and other physical properties~\cite{Millis2022}. 

The parallel SC dome shift with almost the same highest $T_c$ also questions why current IL-nickelates only host much lower $T_c$ than cuprates. One speculation is that, apart from the sample synthesis quality, either one-band or $d$-$s$ type models are not appropriate for describing IL-nickelates because of some missing ingredients proposed in those multi-orbital models~\cite{Hall1,Hall2,GuangMing,Mi2020,MMG2022,dswave,Hanghuireview,Werner2022} that can weaken the SC tendency, or IL-nickelate's $T_c$ still have much room to increase as suggested by the high pressure experiments~\cite{pressure}.

\section{Conclusion}
In conclusion, we systematically investigated the SC properties of a two-orbital $d$-$s$ model to mimic two-dimensional $d$-orbital Hubbard model influenced by a dry $s$ conduction band (dry Fermi sea). To accurately approach to the superconducting (KT transition) $T_{KT}$, we employed the dynamic cluster quantum Monte Carlo focussing on the impact of the $d$-$s$ hybridization in a range of hole doping levels.
Our simulations demonstrated the unusual impact of this additional ``dry'' metallic band, which shift the $d$-wave SC dome of $d$-orbital towards the overdoped regime, namely the SC can be enhanced (suppressed) by a dry conduction band in the overdoped (underdoped) regime. Concurrently, the PG crossover line $T^*(n_d)$ shifts together with $T_{KT}$ such that the underdoped regime hosts PG phenomena that disappears in the overdoped regime.

Based on these observations, we analyzed the $d$-wave effective pairing interaction $\Gamma_d$, dressed and bare pair-field susceptibility $P_{d}, P_{d0}$. It is found that the doping induced SC dome originates from the compromise between the opposite monotonic dependence of $\Gamma_d$ and $P_{d0}$ on $n_d$, namely the same physics as the pure Hubbard model. In contrary, at fixed $n_d$, increasing $d$-$s$ hybridization $V$ does not enhance the intrinsic pairing $P_{d0}$ and the non-monotonic dependent pairing mainly comes from the sole pairing interaction $\Gamma_d$ and also $P_d$. The distinction between two types of non-monotonicity can be reflected in the BSE leading eigenfunctions, for which the comparison with spin susceptibility further indicates the important role of the AF spin fluctuation in mediating the $d$-wave pairing. 

Furthermore, we postulated on the physical origin of the SC dome shift in terms of Kondo-type electron-hole binding induced effective charge carrier density decrease. We emphasize that this binding picture might only apply for the dry $s$ band, whereas too much $s$ electrons would promote longer-range AF correlations between $d$ local moments (via RKKY interaction) that  competes and finally destroys the SC pairing.

These striking properties of composite bilayer system indicates that the seemingly negligible tiny charge carrier density can indeed play a significant role in enhancing or suppressing some physical properties. Besides, the shift of SC dome appears consistent with the higher optimal doping in IL-nickelates compared to cuprates.
Our presented work provides insights on the drastic impact of dry bands as a potential knob of tuning other physical properties of desired systems. In addition, as an extreme variant of PAM, the unusual influence of $d$-$s$ hybridization reported here prompts that in the situations more relevant to heavy fermion physics, there might be more unexpected phenomena to be uncovered.

\section*{Acknowledgments}
This work was supported by National Natural Science Foundation of China (NSFC) Grant No. 12174278, startup fund from Soochow University, and Priority Academic Program Development (PAPD) of Jiangsu Higher Education Institutions. 



\begin{thebibliography}{40}
\bibitem{phonon1} 
Y. He, M. Hashimoto, D. Song, S.-D. Chen, J. He, I. M. Vishik, B. Moritz, D.-H. Lee, N. Nagaosa, J. Zaanen, T. P. Devereaux, Y. Yoshida, H. Eisaki, D. H. Lu, Z.-X. Shen, Science {\bf 362}, 62?65 (2018).

\bibitem{incipient1}
K. Kuroki, T. Higashida, and R. Arita, Phys. Rev. B 72, 212509 (2005).

\bibitem{incipient2}
K. Ochi, H. Tajima, K. Iida, and H. Aoki, Phys. Rev. Research 4, 013032 (2022).

\bibitem{incipient3}
A. Linscheid, S. Maiti, Y. Wang, S. Johnston, and P. J. Hirschfeld, Phys. Rev. Lett. 117, 077003 (2015).

\bibitem{incipient4}
S. Karakuzu, S. Johnston, and T. A. Maier, Phys. Rev. B 104, 245109 (2021).

\bibitem{Werner}
C. Yue, H. Aoki, and P. Werner, arXiv: 2207.12275 (2022).

\bibitem{Kivelson}
S. Kivelson, Phys. B: Condens. Matter 318, 61 (2002).

\bibitem{Berg2008}
E. Berg, D. Orgad, and S. A. Kivelson, Phys. Rev. B 78, 094509 (2008).

\bibitem{Wachtel2012}
G. Wachtel, A. Bar-Yaacov, and D. Orgad, Phys. Rev. B 86, 134531 (2012).

\bibitem{RTS2014}
A. Zujev, R. T. Scalettar, G. G. Batrouni, and P. Sengupta, New J. Phys. 16, 013004 (2014).

\bibitem{Maier2022}
P. M. Dee, S. Johnston, and T. A. Maier, Phys. Rev. B 105, 214502 (2022).

\bibitem{2019Nature}
D. Li, K. Lee, B. Y. Wang, M. Osada, S. Crossley, H. R. Lee, Y. Cui, Y. Hikita, and H. Y. Hwang, Nature (London) 572, 624 (2019).

\bibitem{Pr}
M. Osada, B. Y. Wang, K. Lee, D. Li, and H. Y. Hwang, Phys. Rev. Materials 4, 121801(R) (2020).

\bibitem{La}
M. Osada, B. Y. Wang, B. H. Goodge, S. P. Harvey, K. Lee, D. Li, L. F. Kourkoutis, and H. Y. Hwang, Adv. Mater. 33, 2104083 (2021).

\bibitem{LaSC} 
S. W. Zeng, C. J. Li, L. E. Chow, Y. Cao, Z. T. Zhang, C. S. Tang, X. M. Yin, Z. S. Lim, J. X. Hu, P. Yang, A. Ariando, Science Advances 8, eabl9927 (2022).

\bibitem{Nd6Ni5O12} 
G. A. Pan, D. F. Segedin, H. LaBollita, Q. Song, E. M. Nica et al, Nat. Mater. (2021). 

\bibitem{Junjie_review}
J. Zhang and X. Tao, CrystEngComm,2021,23,3249 (2021).

\bibitem{Botana_review}
A. S. Botana, F. Bernardini, and A. Cano, JETP 159, 711 (2021)

\bibitem{Arita_review}
Y. Nomura and R. Arita, Reports on Progress in Physics 85, 052501 (2022).

\bibitem{Held2022} 
K. Held, L. Si, P. Worm, O. Janson, R. Arita, Z. Zhong, J. M. Tomczak, M. Kitatani, Front. Phys. 9, 810394 (2022).

\bibitem{Hanghuireview} 
H. Chen, A. Hampel, J. Karp, F. Lechermann, and A. Millis, Front. Phys. 10, 835942 (2022).

\bibitem{synthesis} 
K. Lee, B. H. Goodge, D. Li, M. Osada, B. Y. Wang et al, APL Mater. 8, 041107 (2020).

\bibitem{LSAT1} 
X. Ren, Q. Gao, Y. Zhao, H.
Luo, X. Zhou, and Z. Zhu, arXiv:2109.05761 (2021).

\bibitem{LSAT} 
K. Lee, B. Y. Wang, M. Osada, B. H. Goodge, T. C. Wang, et al,
arXiv:2203.02580 (2022).

\bibitem{TPD_ds} 
C. Peng, H.-C. Jiang, B. Moritz, T. P. Devereaux, and C. Jia, arXiv:2110.07593 (2021).

\bibitem{MJ2022}
M. Jiang, arXiv: 2201.12967 (2022).

\bibitem{Oles2022}
T. Plienbumrung, M. Daghofer, M. Schmid, and A. M. Oles, arXiv: 03287 (2022).

\bibitem{Maier2016}
T.A. Maier, P. Staar, V. Mishra, U. Chatterjee, J.C. Campuzano, and D.J. Scalapino, Nat. Commun. 7, 11875 (2016).


\bibitem{Maier2019}
T.A. Maier and D.J. Scalapino, npj Quantum Materials, 4:30 (2019).

\bibitem{dft25}
Y. Gu, S. Zhu, X. Wang, J. Hu, and H. Chen, Communications Physics 3, 84 (2020). 

\bibitem{pressure}
N. N. Wang, M. W. Yang, Z. Yang, K. Y. Chen, H. Zhang, Q. H. Zhang, Z. H. Zhu, Y. Uwatoko, L. Gu, X. L. Dong, K. J. Jin, J. P. Sun, J.-G. Cheng, Nat. Commun. 13, 4367 (2022).

\bibitem{Hettler98} 
M. H. Hettler, A. N. Tahvildar-Zadeh, M. Jarrell, T.
Pruschke, and H. R. Krishnamurthy, Phys. Rev. B {\bf 58}, R7475 (1998).

\bibitem{Maier05} 
T. Maier, M. Jarrell, T. Pruschke, and M. Hettler, Rev. Mod. Phys. {\bf 77}, 1027 (2005).

\bibitem{code}
U. R. Hahner, G. Alvarez, T. A. Maier, R. Solca, P. Staar, M. S. Summers, and T. C. Schulthess, Comput. Phys. Commun. 246, 106709 (2020). The DCA++ code used for this project can be obtained at https://github.com/CompFUSE/DCA.

\bibitem{GullCTAUX} 
E. Gull, P. Werner, O. Parcollet, M. Troyer, Europhys. Lett. {\bf 82}, 57003 (2008).

\bibitem{Scalapino06} 
D. J. Scalapino, Chapter 13 in the "Handbook of High Temperature Superconductivity", J. R. Schrieffer, editor, Springer (2006).

\bibitem{Maier06} 
T. A. Maier, M. S. Jarrell, D. J. Scalapino, Phys. Rev. Lett. {\bf 96}, 047005 (2006).

\bibitem{Qin2022} 
M. Qin, T. Schafer, S. Andergassen, P. Corboz, and E. Gull, Annual Review of Condensed Matter Physics 13, 275 (2022).

\bibitem{Kivelson1995}
V. J. Emery and S. A. Kivelson, Nature 374, 434–437 (1995).

\bibitem{Abrahams1966}
E. Abrahams, and T. Tsuneto, Phys. Rev. 152, 416–432 (1966).

\bibitem{Maier05a}
T. A. Maier, M. Jarrell, T. C. Schulthess, P. R. C. Kent, and J. B. White, Phys. Rev. Lett. 95, 237001 (2005).

\bibitem{Jarrell2013}
K.-S. Chen, Z. Y. Meng, S.-X. Yang, T. Pruschke, J. Moreno, and M. Jarrell, Phys. Rev. B 88, 245110 (2013).

\bibitem{WeiWu2022}
W. Wu, X. Wang, and A.-M. Tremblay, PNAS 119 (13) e2115819119 (2022).

\bibitem{Jarrell01} 
M. Jarrell, T. Maier, C. Huscroft, and S. Moukouri, Phys. Rev. B 64, 195130 (2001).

\bibitem{ScalapinoRMP} 
D. J. Scalapino, Rev. Mod. Phys. 84, 1383 (2012).

\bibitem{Mi20}
M. Jiang, Phys. Rev. B 101, 235124 (2020).

\bibitem{Macridin06} 
A. Macridin, M. Jarrell, T. Maier, P. R. C. Kent, and E. D’Azevedo, Phys. Rev. Lett. 97, 036401 (2006).

\bibitem{Gull2010} 
E. Gull, M. Ferrero, O. Parcollet, A. Georges, and A. J. Millis, Phys. Rev. B 82, 155101 (2010).

\bibitem{Toschi2015} 
O. Gunnarsson, T. Schäfer, J. P. F. LeBlanc, E. Gull, J. Merino, G. Sangiovanni, G. Rohringer, and A. Toschi, Phys. Rev. Lett. 114, 236402 (2015).

\bibitem{MillisGull2022} 
X. Dong, E. Gull, and A. J. Millis, Nat. Phys. (2022). https://doi.org/10.1038/s41567-022-01710-z

\bibitem{Mona2014} 
H. Ebrahimnejad, G. A. Sawatzky, and M. Berciu, Nat. Phys. 10, 951-955 (2014).

\bibitem{dft4}
Y. Nomura, M. Hirayama, T. Tadano, Y. Yoshimoto, K. Nakamura, and R. Arita, Phys. Rev. B 100, 205138 (2019). 

\bibitem{dft101}
M. Kitatani, L. Si, O. Janson, R. Arita, Z. Zhong, and K. Held, npj Quantum Materials 5, 59 (2020).

\bibitem{oneband}
M. Rossi, H. Lu, A. Nag, D. Li, M. Osada, K. Lee, B. Y. Wang, S. Agrestini, M. Garcia-Fernand ez, Y. D. Chuang, Z. X. Shen, H. Y. Hwang, B. Moritz, K.-J. Zhou, T. P. Devereaux, and W. S. Lee, Phys. Rev. B 104, L220505 (2021). 

\bibitem{MMG2022}
M. Jiang, M. Berciu, and G. A. Sawatzky, arXiv:2208.04562 (2022).

\bibitem{Millis2022}
J. Karp, A. Hampel, and A. J. Millis, arXiv:2201.10481 (2022).

\bibitem{Hall1}
D. Li, B. Y. Wang, K. Lee, S. P. Harvey, M. Osada, B. H. Goodge, L. F. Kourkoutis, and H. Y. Hwang, Phys. Rev. Lett. 125, 027001 (2020).

\bibitem{Hall2}
S. Zeng, C. S. Tang, X. Yin, C. Li, M. Li, Z. Huang, J. Hu, W. Liu, G. J. Omar, H. Jani, Z. S. Lim, K. Han, D. Wan, P. Yang, S. J. Pennycook, A. T. S. Wee, and A. Ariando, Phys. Rev. Lett. 125, 147003 (2020).

\bibitem{GuangMing}
G.-M. Zhang, Y.-F. Yang, and F.-C. Zhang, Phys. Rev. B 101, 020501(R) (2020).

\bibitem{dswave}
Q. Gu, Y. Li, S. Wan, H. Li, W. Guo, H. Yang, Q. Li, X. Zhu, X. Pan, Y. Nie, and H.-H. Wen, Nat. Comm. 11, 6027 (2020).

\bibitem{Mi2020}
M. Jiang, M. Berciu, and G. A. Sawatzky, Phys. Rev. Lett. 124, 207004 (2020).

\bibitem{Werner2022}
V. Christiansson, F. Petocchi, and P. Werner, arXiv:2209.04349 (2022).

\end{thebibliography}
\end{document}